\documentclass[twocolumn]{aastex61}
\usepackage{hyperref}
\usepackage{amssymb,amsmath}
\usepackage{url}

\usepackage{graphicx}
\newcommand{\siml}{\lower4pt \hbox{$\buildrel < \over \sim$}}
\newcommand{\simg}{\lower4pt \hbox{$\buildrel > \over \sim$}}

\begin{document}

\title{High-speed Ejecta from the Gamma-ray Binary PSR\,B1259--63/LS\,2883\footnote{
This is a pre-print of an article published in ``Rendiconti Lincei. Scienze Fisiche e Naturali''. The final authenticated version is available online at \url{http://link.springer.com/article/10.1007/s12210-019-00765-0}.
The article is the peer-reviewed version of a contribution selected among those presented at the Conference on Gamma-Ray Astrophysics with the AGILE Satellite held at Accademia Nazionale dei Lincei and Agenzia Spaziale Italiana, Rome on December 11-13, 2017.}
}


\author{George G.\ Pavlov}
\affiliation{Pennsylvania State University, Department of Astronomy \& Astrophysics, 525 Davey Lab., University Park, PA 16802; ggp1@psu.edu}
\author{Jeremy Hare}
\affiliation{George Washington University, Dept.\ of Physics, Washington, DC 20052, USA}
\author{Oleg Kargaltsev}
\affiliation{George Washington University, Dept.\ of Physics, Washington, DC 20052, USA}





\begin{abstract}
Observing the famous high-mass, eccentric X-ray and $\gamma$-ray binary PSR\,B1259--63/LS\,2883 with Chandra, we detected X-ray emitting clumps moving from the binary with speeds of 
$\sim 0.1\,c$, 
possibly with acceleration. 
The clumps are being ejected at least once per binary period, 3.4 yr, 
presumably
 around binary periastrons. The 
power-law spectra of the clumps can be interpreted as
synchrotron emission of relativistic electrons. 
Here we report the results of
8 observations of the clumps in 2011-2017 (two binary cycles)
 and discuss possible interpretations of this unique
phenomenon.
\end{abstract}

\keywords{Pulsars: individual (B1256--63) --- X-rays: binaries}

\section{Introduction}
\label{intro}
\noindent
High-mass $\gamma$-ray binaries (HMGBs)
 consist of a compact object
(a black hole or a neutron star) and a massive B- or O-type companion
(see \citealt{dubus2013}
for a review). If the compact object is a rotation-powered
(non-accreting) pulsar, the collision of the relativistic
pulsar wind with the companion's wind produces
a shock, and the
shocked wind(s) generate synchrotron and inverse Compton (IC) emission from the
radio to TeV $\gamma$-rays \citep{tavani1997}. Observations of this emission allow one to
study the properties of the winds and the binary companions.

Particularly interesting for such studies is the 
 PSR\,B1259--63/LS\,2883 binary (B1259 hereafter),
one of only two  HMGBs in which a 
rotation-powered pulsar has been detected\footnote{
The second 
such binary, PSR\,J2023+4127/MT91\,213, was identified 
by \cite{lyne2015}. Its binary period of $\sim 50$ years is, however, too long to study the orbital dependence in our lifetime.}.
 PSR\,B1259--63 has the
following properties: period $P=48$ ms,
characteristic age $\tau=330$ kyr,
spin-down power $\dot{E}=8.3\times 10^{35}$ erg s$^{-1}$, and distance
$d\approx 2.7$ kpc.
The pulsar's companion, LS\,2883, is a fast-rotating O9.5Ve star, with mass $M_*\sim 30 M_\odot$,
luminosity
$L_*=6.3\times 10^4 L_\odot$,
and an equatorial
disk inclined at $\approx 35^\circ$ to the orbital plane (see \citealt{negueruela2011}).
The results of the VLBI astrometry \citep{miller-jones2018}, combined with
high-precision pulsar timing \citep{shannon2014}, have provided a full orbital
solution for the binary system, including
orbital period $P_{\rm orb}=1236.7$ days,
 eccentricity $e=0.87$,  semimajor axis $a\simeq6$ AU,  inclination angle
$i\simeq 154^\circ$, and longitude of the ascending node $\Omega \simeq 189^\circ$.

Multi-epoch observations of B1259 with numerous X-ray observatories have
revealed binary phase dependences of its flux, photon index, and absorption
column density,
with particularly large variations
within a
(--20,+30) days interval
 around periastron passages \citep{chernyakova2015}.
AGILE,  Fermi and H.E.S.S.\ observations have shown that the
passages of 3 consecutive periastrons
(2010 Dec 14, 2014 May 5, 2017 Sep 22)
were accompanied by strong 
$\gamma$-ray flares, which had no
counterparts at lower energies (see \citealt{tavani2010}, \citealt{abdo2011},
\citealt{hess2013}, \citealt{chernyakova2015}, \citealt{johnson2018}).
Unexpectedly, the flares
occurred within 30--70 day intervals {\it after} peristron passages (10--50
 days  after
the pulsar passed through the stellar disk).
Based on observations of H$\alpha$ lines, whose equivalent widths allow one
to estimate the mass of the equatorial disk of the O star,
it was suggested in \cite{chernyakova2014} that the GeV flares are associated with
 partial destruction of the disk by the passing pulsar.
The flares could be produced by 
IC emission from interactions between the pulsar wind and the stellar
photons \citep{khangulyan2012} or by Doppler boosting of synchrotron radiation from 
the termination shocks in the pulsar wind colliding with the fragments of the destroyed disk \citep{kong2012}.
The competing
models are
still rather uncertain and cannot explain all the observed features.
However, since a fragment of the destroyed disk could be ejected from the binary,
detection and study of such fragments could provide a clue for understanding
the dramatic variations of electromagnetic radiation from B1259 around the
periastron passages, particularly the $\gamma$-ray flares.
In this work we describe the main properties of the X-ray emitting ejecta detected 
in a series of high-resolution  Chandra observations during 2 binary cycles.
Observations in the 2010-2014 binary cycle have been reported by \cite{kargaltsev2014} and \cite{pavlov2015}, while details of the 2014-2017 cycle observations and their analysis 
will be presented in a separate publication.

\begin{figure*}[t]
\includegraphics[width=1.0\hsize]{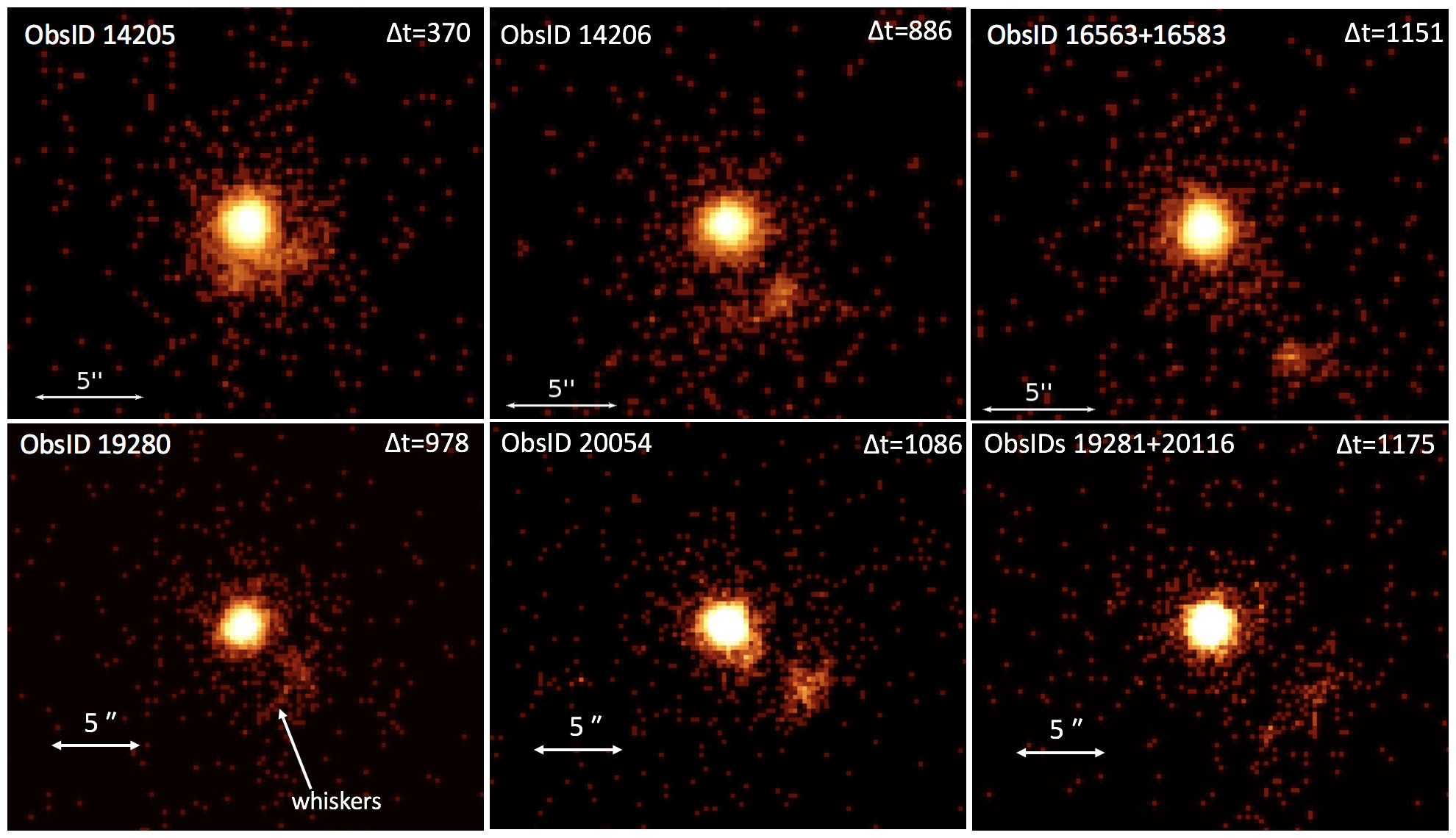}
\caption{
Images of the moving clumps in the binary cycles 2010-2014 (top)
and 2014-2017 (bottom). The $\Delta t$ values show the number of days since
the most recent periastrons. }
\label{clumps}
\end{figure*}

\section{Observations and analysis of fast-moving ejecta}
\label{sec:2}
\begin{figure*}[t]
\includegraphics[width=0.5\hsize]{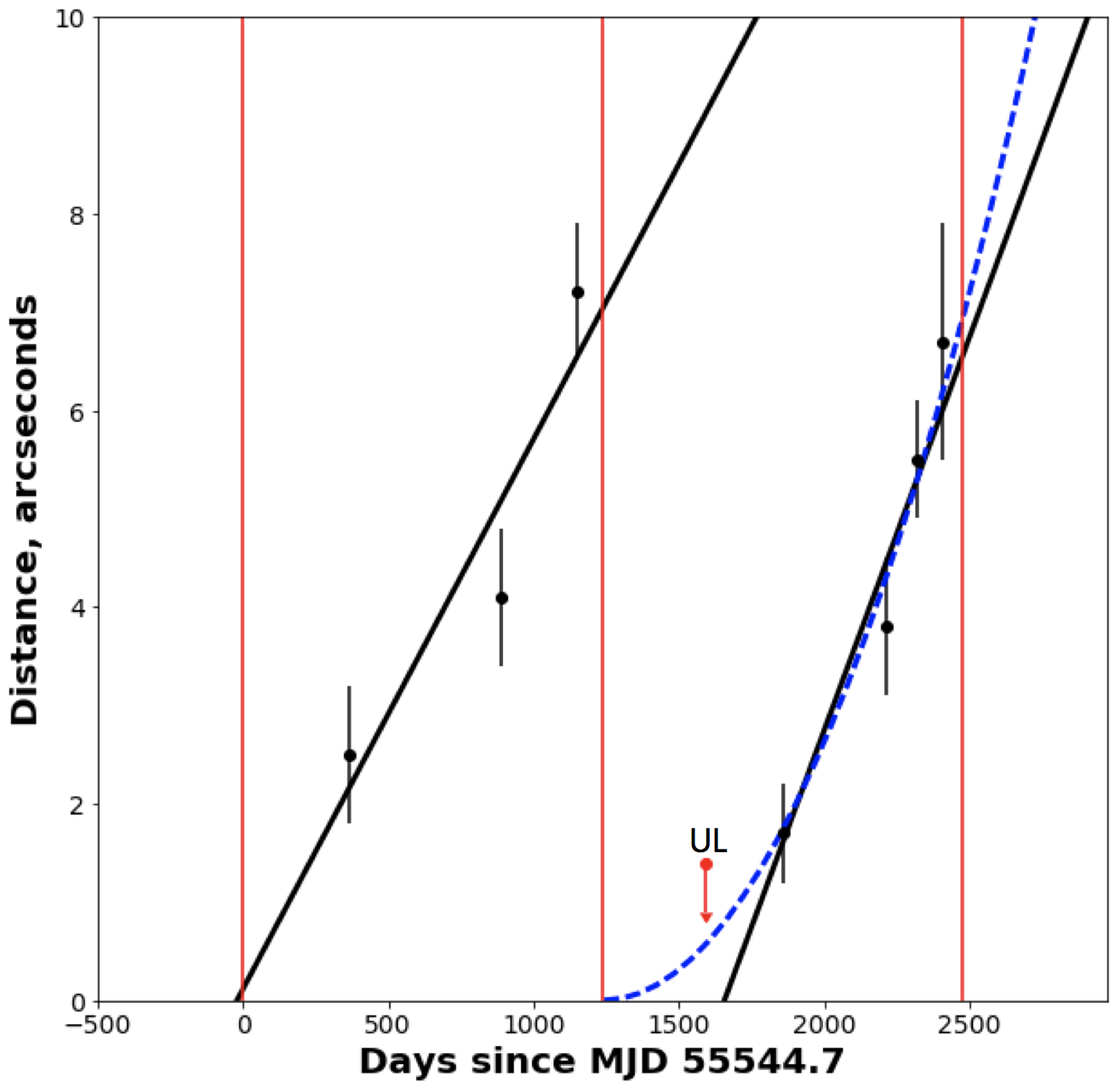}
\includegraphics[width=0.48\hsize]{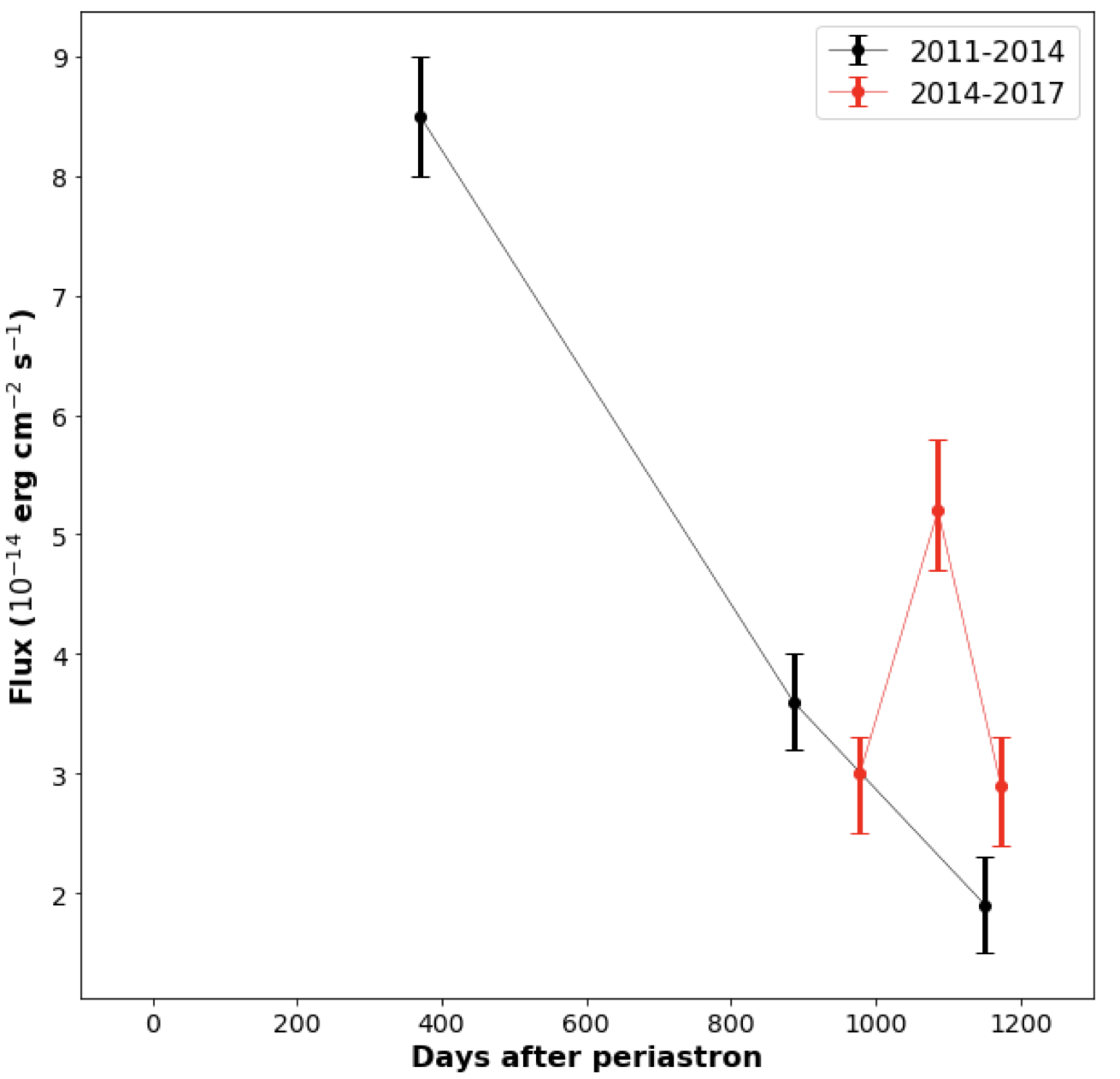}
\caption{
{\em Left:} Separations of the clumps from the binary as a function of time 
after the 2010 and 2014 periastrons. The red vertical lines show the 
periastron epochs. The black solid lines 
show the best 
linear fits,
dashed blue line shows the best quadratic fit for the launch time fixed at the 2014 periastron.
{\em Right:} Clump fluxes in the 0.5--8 keV range
 in units of $10^{-14}$ erg cm$^{-2}$ s$^{-1}$ versus time after periastron 
passage for the 2010-2014 (black) and 2014-2017 (red) orbital cycles.
 }
\label{separations-flux}
\end{figure*}

A southward extension in the B1259 X-ray image was first noticed by \cite{pavlov2011} 
in a 28 ks Chandra ACIS-I observation taken on 2009 May 14, when the binary was close to its apastron ($\Delta t = 667$ days after periastron passage). 
ACIS-I observations in the 2010-2014 cycle were taken at $\Delta t = 370$, 886 and 1151 days after periastron passage. Each of them showed
an extended source
with a characteristic size of $3''$ ($\approx 10^{17}$ cm at $d=2.7$ kpc) at progressively larger distances from the binary (see the upper panels of Fig.\ \ref{clumps}). 
Assuming that it was the same source  moving south-southwest from the binary
(i.e., along the major axis of the binary orbit in the periastron-apastron direction) with a constant velocity,
its projected velocity and launch time were 
 $v_\perp = (0.07\pm0.01)\,c$ and $t_{\rm launch}=-20^{+120}_{-160}$ days, where $t_{\rm launch}=0$ corresponds to the time of periastron passage (see Fig.\ \ref{separations-flux}, left panel).
The 0.5--8 keV luminosity of the source was decreasing from
$8\times 10^{31}$ to 
$2\times 10^{31}$ erg s$^{-1}$ (at $d=2.7$ kpc), but the hard spectrum
of the fading source
(photon index $\Gamma\approx 1.2$)
did not show any softening.
The
source was
interpreted as a clump of stellar disk matter
ejected
from the binary through the interaction of the
pulsar
with the O-star's disk around the 2010 periastron
(possibly at the time of the $\gamma$-ray flare);
its X-ray emission could be due to synchrotron radiation of
the shocked pulsar
wind \citep{pavlov2015}.
Hydrodynamical simulations of clump ejection were presented in \cite{barkov2016}.

To check repeatability of the discovered phenomenon and better understand the clump's properties,
we again observed B1259 in the 2014-2017 binary cycle with the same observational setup and exposure times of about 70 ks.
Five observations 
at $\Delta t = 352$, 620, 978, 1086, and 1175 days from the  2014
periastron 
again showed a clump
apparently moving in the same direction (see the lower panels in Fig.\ \ref{clumps}, where the images from 3 later observations are shown).
However, the clump became discernible only at the second observation.
Moreover, fitting the clump separations from the binary with a linear function of time, we obtained $v_\perp = (0.12\pm 0.02)\,c$ and $t_{\rm launch} = 420^{+78}_{-103}$ days. This places the launch time far after the periastron passage and
after passage through the 
stellar disk (see Fig.\ \ref{separations-flux}, left panel).
To reconcile the time dependence of the clump separation from the binary
with the plausible assumption that it was launched near periastron,
we have to include an acceleration term in the separation versus time fit,
which gives
an acceleration $a=49\pm 2$ cm s$^{-2}=15,400\pm 600$ (km/s)\,yr$^{-1}$ (Fig.\ \ref{separations-flux}, left panel).

 These observations also showed puzzling variations of the clump's
shape and flux at larger $\Delta t$ (see the lower panels in Fig.\ \ref{clumps}),
such as the narrow structure stretched 
perpendicular to the apparent velocity direction (which we call ``whiskers'')
in the third observation
and, in particular,
the unexpected brightening
by a factor of 1.8 in the fourth observation.
Another unexpected feature was evident in the fourth observation:
 an additional clump adjacent to the binary, which disappeared in the fifth
observation.

The evolution of the 0.5--8 keV clump flux values in the two binary cycles is shown in Figure \ref{separations-flux} (right panel). The corresponding clump luminosities vary in the range of (2--$8)\times 10^{31}$ erg s$^{-1}$, or $\sim1\%$--3\% of the 
(variable) binary luminosity in the same energy range.  
The clump spectra could be described by a power-law model in each of the cycles, with photon indices in the range $\Gamma\approx 1.0$--1.6, with a mean value of about 1.4.  

\section{Summary and discussion}
\label{discussion}
\subsection{Summary of observational results}
\label{summary}
Our main findings from the 8 
high-resolution Chandra observations of B1259 can be briefly summarized as follows.
\begin{itemize}
\item B1259 ejects clumps of X-ray emitting matter along the periastron-apastron directions, at least one clump per binary period.
\item A typical
observed size of the resolved clumps
is about $3''$ ($\sim 10^{17}$ cm). The clumps were 
detected up to at least $8''$ from the binary. They showed significant shape
(and size) variations, particularly in the 2014-2017 binary cycle.
\item The apparent velocities $v_\perp$ of the two clumps detected in the 2010-2014 and 2014-2017 
cycles were around 10\% of the speed of light.
\item
Assuming that the clump travels at constant velocity in a given binary cycle, 
its launch time was close to the 
2010 periastron in the 2010-2014 cycle, but it was $\sim 400$ days after the
2014 periastron in the 2014-2017 cycle, in which $v_\perp$ was a factor of 1.7 higher.
\item The assumption that the clumps are launched around the nearest preceding periastrons requires a significant acceleration, $\sim 50$\,cm\,s$^{-2}$, in the 2014-2017 cycle.
\item 
The clump detected in the 2010-2014 cycle faded 
with increasing separation from the binary, while the 2014-2017 cycle clump showed a factor of
1.8 brightening in one of the observations.
\item The clump spectra did not show significant slope variations with time, despite the varying flux.
\item 
An additional clump was detected in the binary vicinity in one of the 2014-2017 observations, but it disappeared 
by the next observation 89 days later.
\end{itemize}

\subsection{Inferences from the clump obervations}

\label{implications}
A detailed analysis of the observational results will be presented elsewhere. Here we only briefly mention the most important inferences.
\paragraph{Repeatability.} In each of the two binary cycles covered by our observations we saw a relatively bright clump moving in the same direction.
However, the different temporal dependences of the clump-binary separation
(Fig.\ \ref{separations-flux}, left) suggest that either the clumps were launched at different binary phases and with different velocities or, if they leave the binary at the same phase (e.g., at periastron), they move with different accelerations. The evolution of the clump shape is noticeably different in the two cycles (Fig.\ \ref{clumps}).  The measured X-ray flux values were similar in the two cycles, but their temporal behavior was different (Fig.\ \ref{separations-flux}, right). The lack of full repeatability 
could be due to the turbulent nature of disk fragmentation and
the interaction of the pulsar wind with the stellar wind  
\citep{barkov2016}.

 \paragraph{The nature of the clump X-ray emission.}
The power-law spectra of the clumps suggest that their emission was produced by 
relativistic electrons, likely
supplied by the pulsar. Possible emission mechanisms are synchrotron radiation in the clump's magnetic field and 
IC scattering of UV photons from the luminous high-mass companion off the relativistic electrons. 
For the IC up-scattering of $\sim 10$ eV photons to X-ray energies, the Lorentz factors of the relativistic electrons should be $\gamma\sim 10$--30. To provide the observed X-ray flux with so low $\gamma$, 
about $10^{50}$ simultaneously emitting electrons is required, 
which is too large a number to be supplied by the pulsar 
or stellar wind. 
Therefore, we conclude that the clump X-rays are emitted by the synchrotron mechanism. For a reasonable magnetic field $B\sim 10$--100 $\mu$G, the Lorentz factors of the X-ray emitting electrons should be in the range $\gamma \sim (0.3$--$3)\times 10^8$. Such parameters are consistent with the lack of synchrotron cooling on a 1000 days timescale.

\paragraph{Clump motion and evolution.}
In both binary cycles the clumps were moving in the periastron-apastron direction.
A lack of any deceleration of the clumps 
implies an exremely low density of the ambient medium (unless the clump is unrealistically heavy) \citep{kargaltsev2014,pavlov2015}. Such a medium could be created by the pulsar wind, which sweeps out the matter of the stellar wind bubble in a cone confined by the shock that separates the colliding stellar and 
pulsar winds \citep{tavani1997,bogovalov2008}. 
Although this cone rotates around the massive star together with the pulsar, 
the cone's axis is close to the periastron-apastron 
direction during most part of the very eccentric orbit, creating a low-density channel, which explains the ejection direction
\citep{barkov2016}. Moreover, the pulsar wind pressure can accelerate the clump, even when it is at large distances from the binary \citep{pavlov2015}. The apparent acceleration in the 2014-2017 cycle can be reached for  clump masses $m_{\rm cl}\siml 10^{23}$\,g.

\paragraph{Conclusion.} The clumps are likely composed of a mixture of stellar matter with the
shocked pulsar wind.
They are formed by the complex interaction of the stellar and pulsar winds.
Formation and initial acceleration of the clumps can be closely connected with the post-periastron $\gamma$-ray flares, but this connection is not certain yet.
The X-ray emitting electrons could be accelerated to $\gamma\sim 10^8$ in shocks or by magnetic field reconnection.
The reconnection in the turbulent clump plasma could be responsible for the observed brightening. 

\begin{acknowledgements}
Support for this work was provided by NASA through Chandra Awards GO5-16065, GO7-18056 and DD7-18088 issued by the Chandra X-ray Center, which is operated by the Smithsonian Astrophysical Observatory for and on behalf of the National Aeronautics Space Administration under contract NAS8-03060.
\end{acknowledgements}

\end{document}